# A SINGLE JOURNAL STUDY: *MALAYSIAN JOURNAL OF COMPUTER SCIENCE*

**A.N. Zainab [1], K.W.U. anyi [1], N.B. Anuar [2]**

[1] Library and Information Unit, Faculty of Computer Science & Information Technology
University of Malaya, 50603 Kuala Lumpur (Malaysia)
[2] Centre of Information Security & Network Research, University of Plymouth, UK.
e-mail:zainab@um.edu.my; nor.jumaat@plymout.ac.uk

*ABSTRACT*

*Single journal studies are reviewed and measures used in the studies are highlighted. The following quantitative measures are used to study 272 articles published in Malaysian Journal of Computer Science, (1) the article productivity of the journal from 1985 to 2007, (2) the observed and expected authorship productivity tested using Lotka's Law of author productivity, identification and listing of core authors; (3) the authorship, co-authorship pattern by authors' country of origin and institutional affiliations; (4) the subject areas of research; (5) the citation analysis of resources referenced as well as the age and half-life of citations; the journals referenced and tested for zonal distribution using Bradford's law of journal scattering; the extent of web citations; and (6) the citations received by articles published in MJCS and impact factor of the journal based on information obtained from Google Scholar, the level of author and journal self-citation.*

## 1. INTRODUCTION

There are an estimated total number of about 189 single journal studies reported in published literature. The first article that traced and reviewed these studies was undertaken by Tiew [1], who retrieved 102 works covering literature published up to the year 1997. The literature found was grouped into four categories; (a) bibliometric studies on single journals (40 items); (b) citation analysis of single journals (45 items); (c) content analyses of single journals (1 item); and other aspects of biblometric study on single journals (6 items). Tiew found that most of the papers were written by authors from the USA (49%), India (20%) and Europe and the rest of the world (31%). Journals in the sciences, technology and medicine (STM) were covered in higher numbers (41%), followed by the library and information sciences (40%) and arts, humanities and social sciences (19%). The majority (84%) of articles were written in the English language.

A more recent review on bibliometrics studies on single journals was carried out by Anyi, Zainab and Anuar [2], covering 82 literatures published from 1997 to 2008. The 82 bibliometric studies on single journals indicated the following situations. Firstly, the number of bibliometric studies on single journals in the sciences and technology remained high with 36% and when this was combined with studies on medical sciences (STM) journals (23%) the proportion increased to 59%. The number of bibliometric studies on journals in the field of library and information science (LIS) was 26% and in the arts, humanities and the social sciences was 15%. Out of the 82 studies, there were 62 unique journal titles as some journals equally in the field of library and information science were revisited in several studies. *JASIST*, *JDoc* and *Scientometrics* were revisited several times during the pre and post 1998 years reflecting their continued influence and importance in sustaining the interests of bibliometrists over the years. Secondly, the majority of journals studied were published in the Asian and African countries (41.4%), followed by those from the USA (30.4%), Europe (18.2%) and the United Kingdom (10.0%). A high number of single journal bibliometrists were Indian and as such there were more contributions from India (28.0%). Out of the 62 unique journal titles studied 30.6% were Indian journal titles. Bibliometric works on single journals began to emerge in other Asian countries such as Malaysia which contributed 6 titles (9.6%). The results indicated that single journal study is of interest to bibliometrists who are fairly distributed worldwide. There is a shift of more contributions from the Asian-African countries instead of the United States as previously indicated by Tiew [1]. Anyi, Zainab and Anuar also found that the journals studied are of some importance in their various fields as reflected by their indexation status. All journals studied are indexed and abstracted by major databases such as *Scopus* or/and *Science Citation Index* or/and the *Social Science Citation Index* as well as major discipline-based indexing databases. Most of the medical and health related journals studies are indexed by Medline. These journals are therefore considered influential or important enough to be studied to identify their publication productivity, authorship and citation patterns, as well the extent of their influence in attracting national and international contributions. Most importantly, the single journal studies have highlighted the variety of bibliometric measures that were used to study the content and format of a journal which subsequently reflected the characteristics of the literature and communication behaviour in the fields they represent.

The studies on single journals highlighed several types of bibliometric measures used. Article productivity or total article output was a common measure used. [3, 4, 5, 6, 7, 8, 9]. This involves counting the number of articles published by issues, volumes, year and sometimes indicating trendlines. This helps infer the publication trend over a period and a journal's influence as a channel for research dissemination amongst authors in a field. Author characteristics was sometimes used as a measure and this refer to author's gender, profession, rank, academic title, author's institutional names and types (academics or professionals). This helps provide a picture or demographic profile of the authors who contribute to a journal and the institutions they are affiliated to [7, 10, 11, 12, 13]. Besides this, authorship productivity is almost always used as a common measure and this involves counting and ranking total authors' contributions and creating a rank list of productive authors. This will reveal the active authors, who are usually also active researchers in the field the journal covers. Authorship productivity is sometimes tested with Lotka's law of authorship distribution. This measure helps identify the key authors in a field and estimate whether the distribution of author productivity are different in the various subject areas [14, 15, 16, 17, 18]. Another measure used is co-authorship pattern, which refers to the types of co-authored works; degree of collaboration; local as well as foreign collaborative activities among authors within and across faculties, universities, institutions and countries. These measures help to highlight the preferred authorship number, the size of the research group in a field and percentage of foreign versus local contributions, which in turn reflect the international acceptance of the journal as a channel to disseminate research [3, 6, 19, 20, 21, 22, 23, 24]. Content analysis is another frequent measure used and this involves analysing the subject areas of articles published in a journal using various classification schemes, keyword analysis, keyword co-occurrence network, article title analysis, word frequency in title, types of research methodology used, types of models, theories and framework used [6, 11, 17, 25, 26, 27, 28, 29, 30, 31]. A number of articles studied the characteristics of the single journal itself to infer quality. Measures observed are the number of pages or length of articles published, journal circulation, journal frequency, analysis of acknowledgement, funding received, article appendices, abstracts, acceptance rate, analysis of indexation and abstraction status, the language of publication; the composition of the editorial and reviewers as well as their gender, qualifications, academic rank, publication productivity and the editorial policy [9, 21, 24, 32, 33]. Citation analysis is also frequently used in single journal studies and this involves analysing the number and distribution of citations referenced per article or volumes over a number of years; the authorship pattern of citations; most referenced author; types of literature cited; the age of cited literature; cited literature's half-life; rank list of core journals cited and testing the spread of journals referenced using Bradford's law of literature dispersion [34]; and the extent and growth of web citations [15, 16, 17, 35, 36, 37, 38, 39]. Finally, another frequent measure used is citation count, which involves counting the number of citations accrued by the articles published in a journal from articles published within the journal itself or elsewhere (sometimes referred to as journal citation image); citation of the journal in other discipline or sub-disciplines (sometimes referred to as journal diffusion rate); geographical location and language distribution of citing articles and self–citation (journal or author self-citation rate) or journal influence or journal diffusion. The citation performance of the journal is sometimes referred to as journal quality and prestige as measured by the journal's impact factor, prestige index, trajectory index, immediacy index, journal attraction power, journal consumption power and discipline contribution score [5, 40, 41, 42, 43]. This paper will apply a selection of the measures mentioned to analyse a single journal in the field of Computer Science and the journal chosen is the *Malaysian Journal of Computer Science*.

## 2.    METHODOLOGY

This study applies selective bibliometric measures to a single Asian journal in the field of computer science, the *Malaysian Journal of Computer Science* (*MJCS*). This journal has been singled out for two reasons. Firstly, it is the only journal in the field of computer science and information technology published in Malaysia that has been in existence for more than ten years. The longevity of this journal indicates that Malaysian and the Asia Pacific authors in this field recognize this journal as an important channel to communicate their research activities and findings. Furthermore, *MJCS* has good indexation records, covered by *Scopus* and the *Science Citation Index* since its 2007 issues as well as *INSPEC* since 1998. The main objective of this paper is to bibliometrically profile *MJCS* using bibliometric measures and this involves; (a) determining the publication productivity of *MJCS* between the years 1985 and 2007; (b) assessing the author productivity of those who had published in *MJCS* and this refers to identifying and ranking the productive authors, testing the author's productivity using Lotka's law of authorship distribution, identifying the core authors and the distribution of author productivity by gender; (c) identifying the authorship pattern in terms of the types of single and joint works, the degree of collaboration among the authors, the country affiliation of authors, the country collaboration, and the institutional affiliation of authors; (d) analysing the content of *MJCS* in terms of subject areas covered, the keyword distribution and the number of words in article titles; (e) analysing the pattern of citations referenced by articles published and this refers to the reference distribution by volume, the types of references cited, the age and half-life of referenced journals, the core journals identified by using Bradford's law of journal distribution; and (f) assessing the pattern of citations received by articles published in *MJCS* in terms of total number of citations received as indicated by *Google Scholar*, the types of citation resources, the author and journal self-citation and the journal's impact factor.

The sample for this study comprises 272 journal articles published in *MJCS* from 1985 to 2007. Access to this journal is obtained from two databases, *EJUM* (*Electronic Journal University of Malaya*), which provides access to full-text from the years 1996 to 2008) available at http://ejum.fsktm.um.edu.my and *MyAIS* (*Malaysian Abstracting and Indexing system*), which provides bibliographic and reference information from 1985 to 2008) available at http://myais.fsktm.um.edu.my. Citations data was obtained from *Google Scholar* since *MJCS* has only just been covered by *Scopus* and *SCI* beginning with its 2007 issues and no citation information can be obtained as yet.

## 3. RESULTS

### 3.1. Article Productivity of MJCS: 1985-2007

The data for this section is collected from *EJUM*'s table of contents from 1996 to 2007 and *MyAIS* from 1985 to 1995. A total of 272 articles were published in the 19 years (1985 to 2007). The number of articles was maximum in 1995 (23 articles) and minimum in 1989 with only 4 articles. The trend lines indicated a continuous decline in articles published per year from 1985 to 1989 with an average productivity of less than 10 articles per year (Figure 1). There were various lapses in publication between 1990 and 1992 and 1994 where no issues were published. The highest number of articles published for a year was in 1995 (23 articles) and ever since the number of articles published per year became constant at between 15 and 20. The trendline (y = 0.656x + 7.754, R² = 0.346) indicates a steady consistent trend in the publication productivity and this trend is expected to remain in the future.

The revival in 1995 may be due to a change in the editorial policy from being national as indicated by these objective statements in 1985 "(a) to assist the academic staff from the University of Malaya and other local Universities in publishing research results and studies in computer science; (b) to provide a medium for discussion and information dissemination on computer applications and advancement of computer science and technology in Malaysia; and (c) to facilitate the planning, evaluation and advancement of computer science education and the application of computers in education in Malaysia" [44], to being more open in 1994 when the statement of objectives was changed to reflect a more international context.

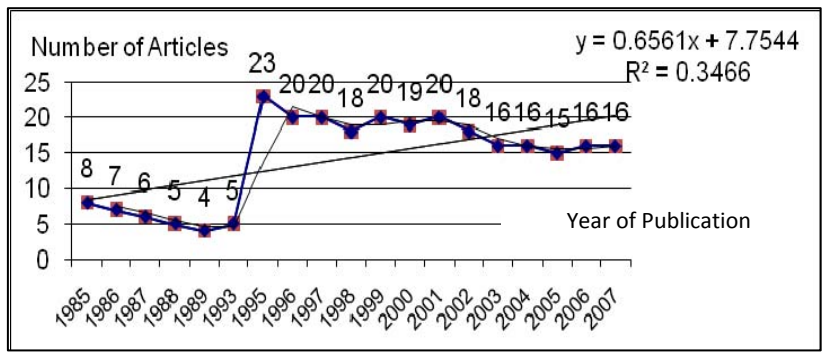

Fig.1. Trend of Articles Productivity in 1985-2007

In 1995, the objectives of *MJCS* was changed; "(a) to promote exchange of information and knowledge in research work, new inventions/developments of computer science and on the use of information technology towards the structuring of an information-rich society; and (b) to assist the academic staff from local and foreign universities, business and industrial sectors, government departments and academic institutions on publishing research results and studies in computer science and information technology through a scholarly publication" [44]. Also, MJCS changed its publication frequency from an annual to a bi-annual since 1995 onwards. This increase in total number of publications may be due to the indexation of *MJCS* by *Inspec*, *Scopu*s and the *ISI*. Keiling and Goncalves [24] noted that the inclusion of the Brazilian journal "*Revista Brasileira de Psiquitria*" in *Medline* (2003) and the *ISI* (2005) had increased the number of submissions received

The sudden increase of articles published in 1995 may also be explained by higher amount of R & D spending on information, computer and communication technologies (ICCT) after 1994 [45]. The percentage of total R&D expenditure on ICCT was increased from 0.6% in 1992 to 9.7% in 1994. This had benefited the computer scientists in Malaysia and the contribution in computer science field had increased by 1.7 times since 1994. According to *MASTIC* (2004), the period of 1996 to 2000 appeared to be the most productive period in Malaysia with highest number of papers produced under the seventh Malaysia Plan.

The publication lapse that occurred between 1990 and 1993 may be explained by Gu and Zainab [46] who reported that during those years there was an up-down trend in research publications productivity by Malaysian authors in computer science and information technology reported in main stream ICT discipline-based databases. The authors found that Malaysian ICT authors used conferences as the primary channel of research communication and this may explain the lack of contributions to journals in those years.

## 3.2. Authorship Productivity Pattern

A total of 424 authors contributed to the 272 articles published in *MJCS* during 1985 to 2007. The articles productivity of authors is shown in Table 1, which indicates that three-quarters (333, 78.5%) of 424 authors had contributed only one article. Only one quarter (91) of authors produced more than two articles between 1985 and 2007 and among the 91 authors, only 13 (2.9%) authors contributed 5 or more articles.

Table 1: Observed and Expected Author Productivity Distribution (n=2.85)

| Number of Articles, x | Number of Authors (observed), y | Observed Percentage (%) | Number of Authors (expected), n=2.85 | Expected Percentage (%) |
|---|---|---|---|---|
| 1 | 333 | 78.50 | 333 | 81.24 |
| 2 | 46 | 10.80 | 46 | 11.22 |
| 3 | 18 | 4.20 | 14 | 3.52 |
| 4 | 14 | 3.30 | 6 | 1.55 |
| 5 | 4 | 0.90 | 3 | 0.82 |
| 6 | 3 | 0.70 | 2 | 0.49 |
| 7 | 2 | 0.50 | 1 | 0.31 |
| 8 | 1 | 0.20 | 1 | 0.21 |
| 9 | 1 | 0.20 | 1 | 0.15 |
| 10 | 0 | 0 | 0 | 0 |
| 11 | 0 | 0 | 0 | 0 |
| 12 | 1 | 0.20 | 0 | 0 |
| 13 | 0 | 0 | 0 | 0 |
| 14 | 0 | 0 | 0 | 0 |
| 15 | 0 | 0 | 0 | 0 |
| 16 | 0 | 0 | 0 | 0 |
| 17 | 0 | 0 | 0 | 0 |
| 18 | 1 | 0.20 | 0 | 0 |
| Total | 424 | 100 | | |

Lotka's law [47] derived the frequency of publication by authors in any given field as $x^n y = c$ where, x is the number of contributors, y is for the number of authors, and c stands for constant. The value of n is 2 based on the results obtained by Lotka's study. Also, Benito et a.l [48] agreed that the exponent n is often fixed at 2 and Sen, Che Azlan & Mohd. Faris [49] found that Lotka's findings may be true to the field of science, where the number of scientific articles published in a year is generally found to be high as compared to other fields of knowledge. The expected percentages calculated based on the Lotka's Law, were compared with the observed percentages of the present study, the result shows that the observed percentages of authors with two or more articles are lower than the expected percentages. This indicates that more authors contributed one articles, whereas a few authors contributed two or more articles. The value of n in this study is 2.85. With the value of n = 2.85, the observed and calculated values were found to be very close (Table 1). This indicates that Lotka's law is applicable to the field of computer science, with a slightly higher n value. This results support a previous study by Liu [50] who observed that the ratios of authors with two or more articles publishing in *JASIST* (field of information science) were lower than expected according to Lotka's Law. Ullah, Butt and Haroon [18] and Patra and Chand [51] also found that Lotka's Law (n=2) does not apply in their research fields (medicine), indicating that authorship distribution pattern may slightly differ in various disciplines

### 3.2.1. Core Authors

There are a total of 424 authors contributing articles between 1985 and 2007 (Table 2). The most productive authors appear to be editorial members of *MJCS*. For an example, Mashkuri Yaacob authored and co-authored 18 articles as well as contributed an average of one article per year based on the 19 issues produced from 1985 to 2007.

Table 2: Ranked List of Most Prolific Contributors

| Group | Author's Name | Number of Articles |
|---|---|---|
| 1 | **Cohort: 1** | 18 |
|   | Mashkuri Yaacob |  |
| 2 | **Cohort: 1** | 12 |
|   | Lee Sai Peck |  |
| 3 | **Cohort: 1** | 9 |
|   | Ling Teck Chaw |  |
| 4 | **Cohort: 1** | 8 |
|   | Mohamed Othman |  |
| 5 | **Cohort: 2** | 7 |
|   | Phang Keat Keong |  |
|   | Zaitun Abu Bakar |  |
| 6 | **Cohort: 3** | 6 |
|   | Elok Robert Tee |  |
|   | Md. Rafiqul Islam |  |
|   | Selvanathan, N. |  |
| 7 | **Cohort: 4** | 5 |
|   | A. K. Ramani |  |
|   | Aziz Deraman |  |
|   | Mustafa Mat Deris |  |
|   | Tengku Mohd. Tengku Sembok |  |
| 8 | **Cohort: 14** | 4 |
|   | Abdul Azim Abdul Ghani |  |
|   | Abdul Rahman Abdullah |  |
|   | Abdullah Mohd Zin |  |
|   | Harihodin Selamat |  |
|   | Kalim Qureshi |  |
|   | Md. Mahbubur Rahim |  |
|   | Mohd. Noor Md. Sap |  |
|   | Ow Siew Hock |  |
|   | Rodina Ahmad |  |
|   | Salina Abdul Samad |  |
|   | Sellappan Palaniappan |  |
|   | V. Prakash |  |
|   | Zaidi Razak |  |
|   | Zarinah Mohd Kasirun |  |
| 9 | **Cohort: 18** | 3 |
|   | Afzaal H. Seyal |  |
|   | Aini Hussain |  |
|   | Ali Mamat |  |
|   | Borhanuddin Mohd Ali |  |
|   | Green, R.J |  |
|   | Jason Teo |  |
|   | Kamal Zuhairi Zamli |  |
|   | Khalid Mohamed Nor |  |
| | Kok Meng Yew | |
| | Liew Chee Sun | |
| | Md. Nasir bin Sulaiman | |
| | Mohamed Daud | |
| | Mohamed Khalil Hani | |
| | Mohd. Noah Abd. Rahman | |
| | Ramlan Mahmod | |
| | S. Selvakennedy | |
| | Wan Ahmad Tajuddin Wan Abdullah | |
| | Zainul Abidin Md. Sharrif | |
| 10 | **Cohort: 46** | 2 |
| 11 | **Cohort: 333** | 1 |

Mashkuri was the founding dean of the Faculty of Computer Science and Information Technology at the University of Malaya in 1994. He had initiated the publication of *MJCS* in 1986 and was the Chief Editor of the journal from 1995 onwards. Similarly, active authors such as Lee Sai Peck, Ling Teck Chaw, Phang Keat Keong and Zaitun Abu Bakar have all served as the executive editors of *MJCS*. It is concluded that most productive authors are senior academician in the field of computer science and information technology in Malaysia and have served as editorial members of *MJCS*. This characteristic is similarly found by other journal studies. According to Young [52], more than 50 percent of the top thirty *Library Quarterly* contributors have served on the editorial board, and a large majority of these contributors were either University of Chicago (which publishes *Library Quarterly*) doctoral graduates or faculty or both. The studies on *Malaysian Journal of Library and Information Science* by Tiew, Abrizah and Kaur [53] and Aryati and Willett [54] indicated likewise that editorial members tended to be the active and core contributors to the journal they were involved in.

### 3.2.2. Author's Gender

Male authors far outnumbered female in contributors to *MJCS*. The data on author's gender was obtained from the biography section of the articles published. Of the 424 authors, 323 (76.18%) were male while 80 (18.87%) were female. There were 21 authors whose gender could not be identified. This result may reflect women's research productivity in journals in the field of computer science and information technology in Malaysia. The predominance of male authors in scholarly journals was also indicated by Prozesky [55] and Sarkar [56].

### 3.3. Authorship Pattern

Out of the 272 articles, two authored articles was the dominant authorship pattern for articles published in *MJCS*. In the order of size of contributions, about 38.6% (105) were two-authored works, followed by 23.5% (64) are single-authored, 23.2% (63) are three-authored, 11.4% (31) are four authored, and 2.9% (8) are 5 authored works. Only one work (0.4%) is authored by 7 people.

When the co-authored works was observed by year the results indicated that joint authored works tended to increase over the recent years (Table 3) especially from 1995 onwards. Five authored articles appeared in 1999 and seven authored article was contributed in 2007 for the first time. The results indicate an inverse situation of increasing joint authored articles and the decline of single authorship over the recent years. This indicated an increase in collaborative research in the field of computer science and information technology.

This finding is similar to the result of quantitative analysis of authorship in *JASIS* by Al-Ghamdi et al. [57] who revealed that shared authorship was more likely between two authors than among three or more authors. Gu and Zainab [58] observed that the collaboration between two authors was the dominant authorship pattern for Malaysian researchers in computer science and information technology. Ullah, Butt and Haroon[18] in their bibliometric study on *Journal of Ayub Medical College*, revealed that two and three author contributions ranked the highest in terms of authorship pattern in medical sciences. Narang [15] also explained that science subjects (except for mathematics) are usually supported by larger research teams and co-authored works sometimes went beyond four researchers.

### 3.3.1. Degree of Author's Collaboration

The degree of collaboration among authors who published in *MJCS* was determined using the formula proposed by Subramanyam [59], C = Nm / (Nm+Ns) (Table 3), where, C = Degree of collaboration; Nm = number of multi authored works; Ns = Number of single authored works.

Table 3: Degree of Collaboration by Year (Subramanyam's Formula)

| Year | No of authors per article | | Degree of Collaboration |
|------|--------|-------|-------------------------|
|      | Single | Multi |                         |
| 1985 | 3      | 5     | 0.63 |
| 1986 | 5      | 2     | 0.29 |
| 1987 | 3      | 3     | 0.50 |
| 1988 | 1      | 4     | 0.80 |
| 1989 | 3      | 1     | 0.25 |
| 1993 | 2      | 3     | 0.60 |
| 1995 | 10     | 13    | 0.57 |
| 1996 | 8      | 12    | 0.60 |
| 1997 | 6      | 14    | 0.70 |
| 1998 | 3      | 15    | 0.83 |
| 1999 | 3      | 17    | 0.85 |
| 2000 | 1      | 18    | 0.95 |
| 2001 | 4      | 16    | 0.80 |
| 2002 | 3      | 15    | 0.83 |
| 2003 | 2      | 14    | 0.88 |
| 2004 | 1      | 15    | 0.94 |
| 2005 | 3      | 12    | 0.80 |
| 2006 | 2      | 14    | 0.88 |
| 2007 | 1      | 15    | 0.94 |
| Total | 64    | 208   | 0.76 |

The degree of collaboration varied from 0.25 to 0.95 and seemed to be inconsistent from 1985 to 1989. From 1993 to 1996, it was almost constant and increased from 1997 to 2007. This result indicated that, like most fields in the sciences, joint authored works in the field of computer science published in *MJCS* is increasing since 1995. Ramesh and Nagaraju [14] similarly found that the degree of collaboration in *International Journal of Tropical Geography* had varied from 0.85 to 0.94. Chaurasia [60] however recorded the degree of collaboration for *Annals of Library and Information Studies* ranged from 0.60 to 0.76. This may indicate that collaborative research is more likely and higher in the field of science and technology.

### 3.3.2. Country Affiliation of Authors

Country contributions indicated that 63.43% of the authors came from Malaysia, followed by United Kingdom (6.25%), Bangladesh (4.40%), Japan (3.70%), and others (22.2%) (Table 4).

Table 4: Authors by Country Affiliations

| Region | Country Affiliation | Number | Percent (%) |
|--------|---------------------|--------|-------------|
| Africa | Morocco(9), Algeria (6), Tunisia (4), Egypt (2) | 21 | 4.86% |
| Australasia | Australia (13), New Zealand (2) | 15 | 3.47% |
| East Asia | Japan (16), Taiwan(10), Korea (8), China (2), Macau (1) | 37 | 8.56% |
| Europe | United Kingdom (27), France(3), Ireland (1), Norway (1) | 32 | 7.41% |
| Middle East | Iran (4), Jordan (3), Saudi Arabia (2), Kuwait (1) | 10 | 2.31% |
| North America | United States (3) | 3 | 0.69% |
| South Asia | Bangladesh (19), India (5), Pakistan (5), Sri Lanka (1) | 30 | 6.94% |
| Southeast Asia | Malaysia (274), Brunei (9) | 283 | 65.51% |
| Unknown | Unknown (1) | 1 | 0.23% |
| Grand Total | | 432 | 100.00% |

This pattern of country affiliation is similarly found by Tiew, Abrizah and Kaur [53] for authors contributing to *Malaysian Journal of library & Information Science* where the top four contributors were geographically affiliated to Malaysia, India, Bangladesh and the United Kingdom. Narang [15] found that the majority of contributors to the *Indian Journal of Pure and Applied Mathematics* were 50.47% from India and 49.52% from foreign countries. Naqvi [61] also recorded 51.39% contributions from the United Kingdom for *Journal of Documentation*. It is thus expected as journals become more visible as a result of their indexation status in established databases they become more international in terms of the number of foreign contributors. The number of articles from the journals country of origin would be reduced as there would be more submissions and acceptance rate from foreign authors. The regional distribution of articles is provided in Table 4, indicating that authors from the Asia-Pacific region were using *MJCS* as the preferred channel for scholarly communicate. Country and cross country collaboration is indicated in Table 5.

Table 5: Country and Cross-country Collaboration

| Types of Contributions | No. of Articles | Percent % | Cumulative No. of Articles | Cumulative Percent |
|---|---|---|---|---|
| Malaysian | 182 | 66.9 | 182 | 66.9 |
| International (non collaboration with other countries | 46 | 16.9 | 228 | 83.8 |
| Malaysian collaboration with international authors | 31 | 11.4 | 259 | 95.2 |
| International (Collaboration between different countries | 12 | 4.4 | 271 | 99.6 |
| Unknown | 1 | 0.4 | 272 | 100.0 |
| Total | 272 | 100.00 | | |

### 3.3.3. Institution Affiliations of Authors

Authors who published in *MJCS* were affiliated to 112 institutions. Most of the authors' were affiliated to institutions of higher learning (100, 89.29%), followed by government organizations (4, 3.57%), private organizations (4, 3.57%) and research organizations (4, 3.57%). This results reveal that the main contributors to *MJCS* are academicians from both Malaysia and abroad. Vijay and Raghavan [32] who analysed the *Journal of Food Science and Technology* had similarly found high contributions from universities, followed by R & D institutions. The high contribution from academicians from universities was also indicated by studies by Biswas, Roy and Sen [9], Narang [15] and Willett [17].

When the total 274 Malaysian contributions (Table 4) were further analysed, it revealed the following authors' university affiliations, University of Malaya (89 articles), followed by Universiti Putra Malaysia (63), Universiti Teknologi Malaysia (28), Universiti Kebanhgsaan Malaysia (22), Universiti Sains Malaysia (21), Multimedia University Malaysia (8), Universiti Malaysia Sabah (7), Mimos (5), Universiti Malaysia Perlis (5), Universiti Utara Malaysia (5), Kolej Universiti Terengganu (3), Universiti Teknologi Mara (3), University College of Science and Technology Malaysia (3), University of Nottingham Malaysia (3) and others (9). This situation indicates that authors affiliated to the institution that publish a journal tended to be the main contributors.

Institutional collaboration per paper indicates 179 (65.1%) out of 272 articles were produce by members within the same faculty of the same university both from Malaysia and abroad (Table 6).

Table 6:  Type of Institutional Collaborations per Article Contributed

| Collaboration Type | No of articles | Percent | Cumulative no of articles | Cumulative Percent |
|---|---|---|---|---|
| Within members of same faculty of same university | 179 | 65.1 | 179 | 65.1 |
| Between faculties within same university | 14 | 5.9 | 193 | 71.0 |
| Between Universities/institutions within same country | 36 | 13.2 | 229 | 84.2 |
| Between universities/institutions of different countries | 43 | 15.8 | 272 | 100 |
| Total | 272 | 100 | | |

This appears to be the norm where the same faculty members tend to work together in collaborations since it could be a supervise/supervisor collaboration within the same faculty. The collaboration across faculties within the same university was low with only 14 articles. Out of the 14 articles the faculties at Universiti Putra Malaysia (UPM) indicated the highest cross faculty collaborative efforts, producing seven articles. Other Malaysian universities indicating cross faculty collaboration are Universiti Sains Malaysia (1), Universiti Teknologi Malaysia (1) and Multimedia University (1). The other 4 articles are the result of cross faculty collaborations from Universities abroad. Cross universities or institutions collaboration within the same country contributed 36 articles out of which 30 articles were cross-University collaborations within Malaysia.

Out of 43 articles derived between universities or institutions of different countries, 31 articles (11.4%) were contributed in collaboration between Malaysians and international authors and 12 articles (0.4%) were from international authors who co-authored with authors from other different countries. Among the 31 articles (collaboration between Malaysian and International authors), 16 articles were contributed by Malaysians who co-authored with authors from the United Kingdom, followed by Malaysia and Bangladesh authors (3 articles). This may be Malaysian authors who have included their supervisor's name from universities abroad or foreign supervisees who has included their Malaysian supervisors' names. A bibliometric study done by MASTIC*S* [62] revealed that in term of collaboration with foreign scientists, Malaysian scientists worked more with those scientists from the United Kingdom.

The study by Omotayo [63] indicated that the journal *Ife Psychologica* also experienced a growth in the ratio of international contributions and is likely to maintain this due to its presence in the Internet. Likewise, *MJCS* is freely accessible in the Internet since 1996 [64] and indexed by *Inspec, Scopus, Science Citation Index* and *Google Scholar*. Hence, this presence has and will attract international contributors to publish their articles in *MJCS* because it is visibility electronically [65].

### 3.4. Subject Areas of Research

In this study the 272 articles were categorized in accordance to the Association for Computing Machinery (ACM) Computing Classification System, 1998 (http://www.acm.org/class/1998) and the distribution is shown in Table 7.

Table 7: Broad Subject Areas Covered by Articles Published in *MJCS*

| Subjects | No. Of Articles | Cumulative no. of Articles | Percent (%) | Cumulative Percent (%) |
|---|---|---|---|---|
| Computing Methodologies | 85 | 85 | 31.25 | 31.25 |
| Software | 69 | 154 | 25.37 | 56.62 |
| Computer Systems Organization | 64 | 218 | 23.53 | 80.15 |
| Information Systems | 31 | 249 | 11.40 | 91.54 |
| Theory of Computation | 14 | 263 | 5.15 | 96.69 |
| Hardware | 7 | 270 | 2.57 | 99.26 |
| Computer Applications | 1 | 271 | 0.37 | 99.63 |
| Data | 1 | 272 | 0.37 | 100.00 |

The keywords provided for each article was also analysed. Keywords is one of the best indicators to understand and grasp the thought content of articles, methodologies used and areas of research addressed [66, 67]. Neural network topped the list as the most frequent keyword mentioned keyword (12 times), followed by expert system (8 times), performance evaluation (6 times), and software engineering (5 times). Eight keywords appeared four times, 16 keywords appeared three times, 62 keywords appeared twice and 741 keywords appeared once. Keyword analysis helps to determine the scope and core content or themes of the research activity of *MJCS*. The result revealed that large number of keywords were assigned to articles indicating that *MJCS* covers a wide scope of research topics. Keywords analysis can be used by future authors to identify areas that are less covered and strategise to fill in the identified gaps.

Analysis of words in article titles also helps to reflect the precision of subject declaration of a journal. This study revealed that the maximum number of words in article titles was 18. The average number of words in article title is 9.22 words. This fits with the word range given by Anthony [68], who conducted a study on the characteristic of article titles in six computer science journals and found that the average title length vary from 8.0 to 9.9 words.

## 3.5. Citation Analysis

There were a total of 4634 citations in the 272 articles published in *MJCS* from volume 1 to 20, 1985 to 2007. The mean citation per year is 243.89. The range of average citations per volume is from 8 to 25 citations with an average of 17. The number of citations referenced in *MJCS* seems to be similar to other science related fields, which indicate that citation pattern is discipline dependent. In the field of science and technology, the citation analysis of *Journal of Natural Rubber Research* indicated an average citations per article of 16.2 [69] and for *Journal of Ayub Medical College*, the average citations was 17.43 [18]. For the *Journal of the Indian Society for Cotton Improvement*, the average number of citations per article was 10.76 [16] and for the *Indian Journal of Pure and Applied Mathematics,* the average citation per paper is almost 11 citations [15]. In the softer fields the number of citations seems to be higher. The bibliometric study of *Family Business Review* revealed that the average reference was 25.6 [33]. According to Naqvi [61], the *Journal of Documentation* has an average of 21 references per article. These studies show that articles in the science and technology fields tended to use fewer citations as compared to those in the arts and social sciences. Also, it may also show that if the journal is more specialized, there could be fewer literatures to be found as references

### 3.5.1. Types of Resources Referenced

Computer science and information technology authors were found to use a wide variety of materials as resource for their research (Table 8). Journal articles remained as the most frequently cited material contributing 1794 (38.71%) citations followed by books (1216, 26.24% citations), and conference papers (1116, 24.08% citations). There was a decline in citing of books after 2000 and an increase in citing conference paper from 1993 onwards. The use of web resources was fairly small but was increasing from 1996 onwards. Perhaps the number of citations of web resources tended to be restricted when submitting to scholarly journals, even though the number of such citations is increasing. Biswas, Roy and Sen [9] found that authors writing in *Economic Botany* started citing web documents from year 1998 with a low citation count (1%). The result of this study seemed to confirm the findings of a study by Leiding [70] who found the usage of journals was still on the rise, even though the incidence of web citations began during the 1996 and 1997 onwards. However, the usage of books started to drop when web citations began to increase.

Table 8: Number and Percentage of Citations Referenced per Article

| Number of Citation Per Article | Number of Articles | Percent (%) |
|---|---|---|
| 0-10 | 88 | 32.35 |
| 11-20 | 109 | 40.07 |
| 21-30 | 47 | 17.28 |
| 31-40 | 19 | 6.99 |
| 41-50 | 4 | 1.47 |
| 51-60 | 2 | 0.74 |
| 61-70 | 2 | 0.74 |
| 71-80 | 0 | 0.00 |
| 81-90 | 0 | 0.00 |
| 91-100 | 1 | 0.37 |
| Total | 272 | 100 |

Furthermore, the types of resources referenced by authors seem to be discipline dependent. Zainab and Goi [71] observed that humanities researchers were more monograph oriented than other disciplines. Their findings showed that humanities researchers cited more books (52%) followed by journal articles (23.55%). Tiew [10] observed the same trend in the type of materials used by historians publishing in *Journal of the Malaysian Branch of the Royal Asiatic Society (JMBRAS)* who cited 37.7% to books and 20.44% to journal articles. In the field of medical science, Ullah, Butt and Haroon [16] found out that 77.94% citations were from journals while 10.28% from books. *Dixit and Katare* (2007) who studied a journal on cotton research also obtained similar results where 71.93% of their authors cited journals, followed by conference proceedings (9.14%) and books (7.38%). Borkenhagen et al., [3] who examined the international journal *Psychotherapy Research* found that articles referenced more journals (58%) and books (36%). Cited items in *American Journal of Veterinary Research* were journal (88.8%) and books (9.8%) [39]. In summary, citations referenced in *MJCS* seem to conform to the citation practices of articles in the field of science and technology where the use of journals surpass the use of other types of resources.

### 3.5.2. Age of Citations and Publication Half-Life

The analysis of the age of the citations helps to determine the useful life of information resources used in any fields of knowledge. It is also used by academic librarians to maintain or discard monographs and serials in the library which would be no longer needed by researchers. Burton and Kebler [72] used the term "half-life" as the time during which one-half of all the currently active (cited) literature is published. When this is applied to journals as a resource, it refers to the number of journal publication years from the current year going back, which accounts for 50% of the total number of references given by the citing journal. It can be used to measure how fast literature in a certain field becomes obsolete. Table 9 shows that most of the cited literatures used were about four to six years old, ranging from 402 to 440 citations. The maximum citations were 5 years old. This indicates that literatures published between four and six years were relevant to researchers in the field of computer science.

When the data from Table 9 was plotted on a graph with X axis as year bands and Y axis as the cumulative number of citations the calculated half-life of the citations is about 6 years. On the average the half-life of total citations used by authors published in *MJCS* is about 6 years which means that articles older than this will less likely be used by researchers. This confirmed that currency is important in the field of computer science and information technology. Age of literature use seems to be discipline dependent. Wertheimer [73] found that currency was less important in the humanities because the average age of literature use in the humanities was 40 years for monographs and 32 years for journals. Haridasan and Kulshrestha [74] found that the cited half-life of the literature in information science journal *Knowledge Organization* is 14 years. The mean age of citations for two economics journals, namely *Pakistani Development Review* and *Pakistan Economic and Social Review* was 9.16 and 12.96 respectively [75].

Table 9: Age Distribution of Cited Literature

| Age of Cited Literature (year) | Total | Percent (%) |
|---|---|---|
| Up to 1 | 374 | 8.07 |
| 2 | 336 | 7.25 |
| 3 | 376 | 8.11 |
| 4 | 402 | 8.68 |
| 5 | 440 | 9.50 |
| 6 | 413 | 8.91 |
| 7 | 313 | 6.75 |
| 8 | 312 | 6.73 |
| 9 | 219 | 4.73 |
| 10 | 206 | 4.45 |
| 11 | 150 | 3.24 |
| 12 | 112 | 2.42 |
| 13 | 140 | 3.02 |
| 14 | 87 | 1.88 |
| 15 | 84 | 1.81 |
| 16 | 60 | 1.29 |
| 17 | 62 | 1.34 |
| 18 | 48 | 1.04 |
| 19 | 46 | 0.99 |
| 20 | 29 | 0.63 |
| 21 or more | 235 | 5.07 |
| Not Available | 190 | 4.10 |

### 3.5.3. Journal Titles Referenced in Articles

The spread of articles in journals cited in *MJCS* followed a pattern predicted by Bradford's Law of Scattering, indicating three productivity zones [34], where the number of journals published increased from one zone to the next according to the expression 1, n, n2,.. The first zone is the group of journals concentrated in the central core and which exert a certain influence in a field over the other journals [39, 48] compared to the other two zones. There were 628 journal titles referenced in MJCS that contributed a total of 1092 citations. In summary, the three zones comprised:

    a.   Zone 1 (most productive): Top 6 (0.95%) journals produced 364 (33.33%) citations.
    b.   Zone 2 (moderate productive): Next 41 (6.52%) journals produced 381 (34.89%) citations.
    c.   Zone 3 (low productive): Next 581 (92.51%) journals produced 347 (31.77%) citations.

Hence, the ratio of the number of journals in the three zones is 6:41:581. For this study, the estimated value n = 6.8 is found. *Bradford* (1985) suggests the value of n to be 5 as a representative number on the data for his studies on applied geophysics and lubrication. The result of this study does partially comply with the Bradford's Law. The top ten groupings of journal titles used by *MJCS* authors were *IEEE Transaction on Software Engineering* (86 citations), *Communication of the ACM* (59), *IEEE Software* (39), *IEEE Transaction on Computer* (37), *IEEE Transaction on Pattern Analysis and Machine Intelligence* (35), *IEEE Transaction on System, Man, and Cybernetics* (33), *Computer Journal* (32), *IEEE Computer* (29), *Fuzzy Sets and Systems* (26), and *Pattern Recognition* (25).

### 3.5.4. Website Citations

In this study, websites were differentiated from electronic journals. Articles in *MJCS* cited a total of 245 websites. The web resources began to be cited from 1996 onwards and the number has increased. The authors had cited company website (.com) the most (122, 49.80%), followed by education (.edu and .ac) institution websites (64, 26.12%), organization (.org) websites (33, 13.47%), government (.gov) websites (9, 3.67%), network (.net) websites (9, 3.67%), and military (.mil) website (1, 0.41%). Kushkowski [76] reported on the characteristics of web citations found in theses submitted by economics graduate students at Iowa State and Virginia Tech from 1997 to 2003. The result showed that the web citations made up a small portion of the total citation (2.2% and 5.4% respectively). Vaughan and Shaw [77] conducted a study on 46 library and information science journals where they categorized the web citations by types of domains. The result was 57% are from education sites (.edu), and there were fewer web citation counts from company websites (.com). The most common type of citing domain in their study was .edu, followed by .org and .com. However, the findings of the current study differ since almost half of the total web citations were from company website. This may be because the authors tend to refer to certain products which were related to their research. Also, the authors were citing web white papers and reports published by various educational institutions.

### 3.6.   Total Citations Received by MJCS Articles

*Google Scholar* indexes all the articles published in the electronic version of *MJCS,* which parks itself in a journal host *EJUM* (*Electronic Journal University of Malaya*) and available on open access. The citations information given by *Google Scholar* was used to determine the total citations received by *MJCS* articles (Table 10). There was a total of 45 *MJCS* articles published from 1996 to 2005 which had received 80 citations from various literatures published between 1996 and 2008 and reported in *Google Scholar*. Articles published in *MJCS* in 1996 received the highest number of citations (21 citations), followed by articles published in 1997 (18 citations) and, in 2001 and 1999 (9 citations each). This pattern indicates that older articles published in *MJCS* were being cited more. Also, the articles published recently are getting early citations, in some cases within a year. This may be because the journal is made available on open access over the web. Normally, citations will be received after 2 to 3 years for an article. Moed et al. [78] revealed that the number of citations per publication will reach the peak in the second year after their publication date, after which the citations would decline rapidly. The results from this study agree with Moed et al.'s finding that the citations received by *MJCS* were mostly within the first 2 years after their publication. Among the 80 citations, 29 came from Malaysia and 44 citations from foreign countries. This may indicate that *MJCS* is well received by international authors especially among the Asia-Pacific authors. Out of 29 citations from Malaysia, 21 were author self citations (72.41%). On the contrary, citations form foreign countries revealed only 12 author self citations (27.27). In brief, Malaysian authors show preference to cite their own publications in their articles.

Table 10: Total Citations Received by *MJCS* Articles

| Publication Year | Total Articles being cited | Percent (%) | Cumulative Percent (%) | Citation in Year (Number of times) | Total Citations | Percent (%) | Cumulative Percent (%) |
|---|---|---|---|---|---|---|---|
| 2005 | 2 | 4.4 | 4.4 | 2007 (2) | 2 | 2.41 | 2.41 |
| 2004 | 5 | 11.1 | 15.6 | 2005 (3) 2006 (1) 2007 (1) 2008 (1) | 6 | 7.23 | 9.64 |
| 2003 | 1 | 2.2 | 17.8 | 2004 (1) | 1 | 1.20 | 10.84 |
| 2002 | 2 | 4.4 | 22.2 | 2003 (1) 2004 (1) | 2 | 2.41 | 13.25 |
| 2001 | 5 | 11.1 | 33.3 | 2002 (4) 2005 (2) 2006 (2) 2007 (1) | 9 | 13.25 | 26.51 |
| 2000 | 4 | 8.9 | 42.2 | 2000 (1) 2001 (1) 2003 (1) 2005 (1) 2006 (1) | 5 | 6.02 | 32.53 |
| 1999 | 6 | 13.3 | 55.6 | 2000 (3) 2003 (1) 2004 (1) 2005 (1) 2006 (1) 2007 (2) | 9 | 10.84 | 43.37 |
| 1998 | 6 | 13.3 | 68.9 | 1998 (1) 1999 (4) 2002 (1) 2006 (1) | 7 | 8.43 | 51.81 |
| 1997 | 8 | 17.8 | 86.7 | 1997 (2) 1998 (2) 1999 (3) 2000 (1) 2001 (1) 2003 (2) 2004 (2) 2005 (3) 2006 (1) 2007 (1) | 18 | 22.89 | 74.70 |
| 1996 | 6 | 13.3 | 100.0 | 1996 (1) 1997 (2) 1998 (2) 1999 (3) 2000 (2) 2001 (2) 2003 (1) 2004 (2) 2006 (6) | 21 | 25.30 | 100.00 |
| **Total** | **45** | **100** | | | **80** | **100** | |

### 3.6.1. Journal Self-Citations

The findings of this study show the degree of journal self-citation is low (7.3%). According to Omotayo [63], low journal self-citation (1.2%) indicated that the contributors may have derived their supporting literature from sources other than the journal itself. High journal self-citation may show the level of comparative acceptance of the journal as a reliable medium for authors to support their own findings. Hyland [79] analyzed 240 research articles from ten journals in eight disciplines and observed that journal self-citations was a significant means of promoting scholarly reputation and gaining professional credit for one's research. Journal self-citation may be discipline dependent. Hyland's study revealed that articles in the hard sciences such as mechanical engineering, electrical engineering, marketing, philosophy, sociology, applied linguistics, physics, and microbiology had higher journal self-citations (12%), compared with only 4% in the soft fields (marketing, philosophy, sociology and applied linguistics). Tagliacozzo [80] observed that areas in the social sciences indicated 5% journal self-citations compared to between 10% and 20% in the sciences. The result indicates for *MJCS*, journal self-citations is still relatively low, which is surprising since it has been published more than 10 years and one would expect that at least Asia Pacific authors would find it a useful resource to support their research needs.

### 3.6.2. Journal Impact Factor

The impact of a journal depends on how often articles in that journal are cited by other academic publications [81]. The annual *Journal Citation Reports (JCR)* defined impact factor as a ratio between citations and recent citeable items published. Thus, the impact factor of a journal is calculated by dividing the number of current year citations to the source items published in that journal during the previous two years. Table 11 shows the calculated Journal Impact Factor (JIF) for *MJCS* based on the 80 citation data obtained from *Google Scholar*.

Table 11: Journal Impact Factors by Year for *MJCS* based on *Google Scholar*

| Year | Journal Impact Factor | 5 Years Impact Factor | Impact Factor revised to exclude self-citations |
|---|---|---|---|
| 2007 | 0.0645 | 0.0494 | 0.0323 |
| 2006 | 0.0323 | 0.0353 | 0.0323 |
| 2005 | 0.0938 | 0.0787 | 0 |
| 2004 | 0.0294 | 0.0215 | 0.0294 |
| 2003 | 0.0263 | 0.0316 | 0 |
| 2002 | 0.1026 | 0.0515 | 0.0513 |
| 2001 | 0.0256 | 0.0412 | 0 |
| 2000 | 0.1053 | 0.0693 | 0.0789 |
| 1999 | 0.1842 | 0.1235 | 0.0526 |
| 1998 | 0.1250 | 0.0735 | 0.0750 |
| 1997 | 0.0930 | 0.0833 | 0.0698 |
| 1996 | 0.0435 | 0.0357 | 0 |
| 1985-1995 | 0 | 0 | 0 |
| **Average** | **0.0402** | **0.0301** | **0.0183** |

The results of the Journal Impact Factor indicated that articles published in issues from 1996 onwards seem to attract more citations. However, we realize that the citation information is limited in *Google Scholar* as it captures information only from articles openly available on the web. This means that citations of articles published in *MJCS* in non open-access journals, theses and non open-access conference proceedings are not captured. Even though the citation count is low, the growing number of citations received indicates the growing importance of this journal. Price and Jeffrey [82] suggested that publishing more high-quality review articles instead of original research articles could help increase a journal's Impact Factor. The study by Buznik, Zibareva and Piottukh-peletskii [5] on *Journal of Structural Chemistry (JSC)* supported Price and Jeffrey's [82] suggestion. Despite its low publication, the *JSC*'s review articles were cited more widely than its research articles. This gave *JSC* a higher Impact Factor. Perhaps *MJCS* should consider this option to increase its Impact Factor count.

## 4.    Conclusion

This single journal study helps characterize *MJCS* in various lights [68]. When a single journal is studied bibliometrically, it created a portrait of the journal, provided an insight that is beyond the superficial. It helped indicate the quality, maturity and productivity of the journal. It informs about the research orientation that the journal supports to disseminate and its influence on author's choice as a channel to communicate or retrieve information for their research needs. It helps indicate the importance or significance of the journal in a field and somehow reflect the activity of research in the field. Studying the single journal allows one to determine its quality in terms of indexation and impact as well as how it supports joint or collaborative works either within faculty members or cross faculties within the same or different universities at country or international levels. Nebelong-Bonnevie and Frandsen [83] indicated that single journal studies provided a detailed multi-faceted picture of the characteristics of a single journal. This is what we have attempted to do with *MJCS*. The assessment tool used for single journal studies is almost always bibliometric measures and we have utilized selected measures and applied it to study the *Malaysian Journal of Computer Science* (*MJCS*). In summary, analyses of articles published in and citations received by MJCS indicate that (a) a ratio of 60:40 (Foreign:Malaysian) aricles should ideally be maintained to infer its international characteristics; (b) Joint authored articles should be encouraged especially across universities in Malaysia and abroad; (c) include thorough and quality review articles to help boost future citations to the journal; and (d) perhaps increase the frequency of issues to stimulate and encourage higher number of quality submissions.

This study also revealed the variety of bibliometric measures that can be used to understand the characteristics or portrait of a journal which reflect the characteristics of the literature and communication behaviour in the fields they represent. All data are obtained from the bibliographic information provided by articles published in the journal and accompanying author and journal information. Single journal studies have highlighted these bibliometric measures to characterize a journal.

  a)  Article productivity - Number of articles published by issues, volumes and year with trendlines. This helps infer the publication trend over a period and its influence as a channel for research dissemination amongst authors in the field.

b) Author characteristics - Author's gender, profession, rank, academic title; author's geographical affiliations by institutional names and types of institutions (academic, professionals; authors location by region, or country. This helps provide a picture or profile of the authors, the institutions or country they are affiliated to and the degree of collaboration that exists.

c) Author's productivity - Rank list of core and active authors; authorship productivity pattern may be tested with laws of authorship distribution such as Lotka's law. This helps to identify the key authors in a field and estimate whether the distribution of author productivity are different in the various subject area.

d) Co-authorship pattern - Types of co-authored works; degree of collaboration; local and foreign collaboration activities among authors by country and institution, internationalization status of the journal [84]. This helps to highlight the preferred authorship number, the size of the research group in a field and percentage of foreign versus local contributions.

e) Content analysis - Subject areas of articles, keyword analysis, keyword co-occurrence network; article title analysis, number of words, punctuation usage, word frequency and preposition usage [85]; number of pages per article; journal circulation; journal frequency; types of research methodology used; types of models, theories and framework used; analysis of acknowledgement [86]; analysis of funding received; characteristics of the editorial board [87]; editorial policy; analysis of article appendices; analysis of article abstracts.; the acceptance rate, analysis of indexation, abstraction information and the language of publication.

f) Citation analysis - Number and distribution of citations per article, volumes and years; authorship pattern of citations; author co-citation analysis network, most cited author, types of literature cited; age of cited literature; cited literature's half-life, rank list of core journals using Bradford's law, extent and growth of web citations, journal citation image, analysis of citations to the journal, journal influence and diffusion in other subject areas, geographical location and language distribution of cited literature; journal self-citation; author self-citations; journal performance, quality and prestige as measured by journal impact factor, prestige index, trajectory index, immediacy index, journal attraction power, journal consumption power and discipline contribution score.

## Acknowledgement

This study was made possible as a result of funding obtained from the Ministry of Higher Education, Malaysia and the University of Malaya, FP0561/2006A and FS250/2008B, which resulted in the development of MyaIS, the information system that provided the data for this study.